\documentclass[aps,prl,floatfix,superscriptaddress,preprintnumbers,amsmath,amssymb,showpacs,showkeys]{revtex4}
\usepackage{amssymb}
\usepackage{amsmath}
\usepackage{amsfonts}
\usepackage{epsfig}
\usepackage{graphicx}
\usepackage{tabularx}
\usepackage{color}
\newcommand{\seq}{\begin{subequations}}
\newcommand{\sen}{\end{subequatons}}
\newcommand{\eq}{\begin{eqnarray}}
\newcommand{\en}{\end{eqnarray}}            

\def\shiftdown#1{#1\llap{\lower.04ex\hbox{#1}}}

\begin{document}

\title{Nuclear physics in soft-wall AdS/QCD: \\
Deuteron electromagnetic form factors}

\author{Thomas Gutsche}
\affiliation{Institut f\"ur Theoretische Physik,
Universit\"at T\"ubingen, \\
Kepler Center for Astro and Particle Physics,  
Auf der Morgenstelle 14, D-72076 T\"ubingen, Germany}
\author{Valery E. Lyubovitskij} 
\affiliation{Institut f\"ur Theoretische Physik,
Universit\"at T\"ubingen, \\
Kepler Center for Astro and Particle Physics, 
Auf der Morgenstelle 14, D-72076 T\"ubingen, Germany}
\affiliation{Department of Physics, Tomsk State University,  
634050 Tomsk, Russia} 
\affiliation{Mathematical Physics Department, 
Tomsk Polytechnic University, 
Lenin Avenue 30, 634050 Tomsk, Russia} 
\author{Ivan Schmidt}
\affiliation{Departamento de F\'\i sica y Centro Cient\'\i
fico Tecnol\'ogico de Valpara\'\i so (CCTVal), Universidad T\'ecnica
Federico Santa Mar\'\i a, Casilla 110-V, Valpara\'\i so, Chile}
\author{Alfredo Vega}
\affiliation{Instituto de F\'isica y Astronom\'ia, 
Universidad de Valpara\'iso,  
Avenida Gran Breta\~na 1111, Valpara\'iso, Chile}
\affiliation{Centro de Astrof\'isica de Valpara\'iso, 
Avenida Gran Breta\~na 1111, Valpara\'iso, Chile} 

\vspace*{.2cm}

\date{\today}

\begin{abstract}

We present a high-quality description of the deuteron electromagnetic 
form factors in a soft-wall AdS/QCD approach. 
We first propose an effective action 
describing the dynamics of the deuteron in the presence of
an external vector field. Based on this action the deuteron
electromagnetic form factors are calculated, displaying the
correct $1/Q^{10}$ power scaling for large $Q^2$ values. This finding is
consistent with quark counting rules and the earlier observation that this
result holds in confining gauge/gravity duals. 
The $Q^2$ dependence of the deuteron form factors is defined 
by a single and universal scale parameter $\kappa$, which is fixed from data.

\end{abstract}
\vspace*{.5cm}
\date{\today}



\maketitle

The experimental and theoretical study of the deuteron is one of the 
main focuses of hadronic physics during the last decades (for detailed 
reviews see e.g. Refs.~\cite{Garcon:2001sz}-\cite{Holt:2012gg}). 
Many theoretical approaches have been applied to the problem 
of the deuteron form factors: perturbative QCD, chiral effective and 
phenomenological approaches, potential and quark models 
(see e.g. Refs.~\cite{Garcon:2001sz}-\cite{Ito:1989nv}). 
For example, in potential models the nonrelativistic impuls 
approximation~\cite{De Forest:1966dn,Donnelly:1975ze} was used. 
It leads to deuteron form factors factorized in
terms of the isoscalar combinations of the nucleon form factors.
These approaches are able to describe data up to 0.5 GeV$^2$, but
deviate from data for higher $Q^2$ and are not consistent with 
quark counting rules. To include relativistic effects different types
of relativistic nuclear models have been developed.
One possibility is based on taking into account relativistic
corrections in a $v/c$ expansion of the
nonrelativistic current (leading to so-called two-body interaction
current diagrams)~\cite{Friar:1975zza,Mathiot:1989vw,Schiavilla:1990ug}. 
Such approaches are limited in their validity of the description of data
up to 1-2 GeV$^2$. There is a group of models 
based on relativistic Hamiltonian constraint dynamics, which uses
certain phenomenological potentials (Argonne, Nijmegen, etc.) and
three forms of quantization procedures (point, instant or front form)
[see e.g. Refs.~\cite{Lev:1999me,Allen:2000xy}]. 
Field-theoretical methods formulated in terms of
hadronic (mesons, nucleons, $\Delta$-isobars) degrees of freedom are used in
a wide range of approaches. These include models
based on the  solution of a quasipotential~\cite{Blankenbecler:1965gx} 
or on Bethe-Salpeter~\cite{Buck:1979ff} equations. These methods also include
field theories quantized on the light 
cone~\cite{Carbonell:1998rj,Karmanov:1991fv},  
phenomenological Lagrangian approaches~\cite{Dong:2008mt}                     
and effective field theories treating the long-range dynamics explicitly
while parametrizing the short-distance effects by contact interactions
[for recent applications
to deuteron form factors see e.g. Refs.~\cite{Kolling:2012cs}].          
Another class of approaches supposes to treat the deuteron in terms of
fundamental degrees of freedom - quarks and gluons: nonrelativistic
quark models~\cite{Buchmann:1989ua,Ito:1989nv} and 
perturbative QCD~\cite{Brodsky:1983vf}.  
The analysis of Ref.~\cite{Brodsky:1983vf} results in a prediction for
the asymptotic large-momentum-transfer behavior of the 
deuteron form factors and the form of the deuteron distribution 
amplitude at short distances. 
Later on in Ref.~\cite{Polchinski:2001tt} it was shown that 
field theories based on gauge/gravity duality, 
as proposed in Refs.~\cite{Maldacena:1997re},
produce the correct power scaling of hadronic form factors 
at large momentum transfer. This finding is consistent 
with the quark counting rules. 

The main advantage of our approach is that it gives a description of
the deuteron electromagnetic (EM) form factors in terms of a single
dimensional parameter $\kappa$ with the correct power scaling $1/Q^{10}$
at large $Q^2$ as predicted by perturbative QCD. Our approach is
constructed as a holographic dual to perturbative QCD.
It gives a good starting point for studying more complicated many-body
nuclear systems.

Encouraged by this property we focus on
soft-wall anti-de Sitter/quantum chromodynamics 
(AdS/QCD)~\cite{AdSQCD_int1,AdSQCD_int2,AdSQCD_int3,AdSQCD}.
It is a version of a bottom-up approach based on the correspondence 
of string theory in AdS space and conformal field theory (CFT) 
in physical space-time. 
The formalism in soft-wall AdS/QCD is based on an effective 
action involving five-dimensional fields propagating in AdS space, which are 
dual to the deuteron and the electromagnetic field. We apply our formalism 
to the calculation of the EM form factors of the deuteron.
The deuteron itself is simply considered as a proton-neutron bound state. 

Note that there are already some applications of AdS/CFT and AdS/QCD to 
different problems in nuclear physics: baryon matter at finite 
temperature and baryon number density~\cite{Bergman:2007wp}, 
cold nuclear matter~\cite{Rozali:2007rx}, 
baryon-charge chemical potential~\cite{Nakamura:2007nx}, 
$\rho$ meson condensation at finite isospin chemical 
potential~\cite{Aharony:2007uu}, 
holographic nuclear matter~\cite{Kim:2007xi}, 
heavy atomic nuclei\cite{Hashimoto:2008jq}, 
nuclear matter to strange matter transition~\cite{Kim:2009ey}, 
self-bound dense objects\cite{Kim:2010an}, and 
mean-field theory for baryon many-body systems~\cite{Harada:2011aa} 
(for reviews see e.g. Refs.\cite{Kim:2011ey,Kim:2012ey,%
Aoki:2012th,Pahlavani:2014dma}). 

Our approach is based on an effective action, which in terms of the
AdS fields $d^M(x,z)$ and $V^M(x,z)$, is dual to 
the Fock component contributing to the deuteron with twist 
$\tau = 6$, and the electromagnetic field, respectively, is given by 
\eq\label{Eff_action} 
S &=& \int d^4xdz \, e^{-\varphi(z)} \, 
\biggl[ - \frac{1}{4} F_{MN}(x,z) F^{MN}(x,z) 
- D^M d^\dagger_{N}(x,z) D_M d^N(x,z)  
- i c_2 F^{MN}(x,z) d^\dagger_{M}(x,z) d_{N}(x,z)
\nonumber\\ 
&+& \frac{c_3}{4M_d^2} \, e^{2A(z)} \, \partial^M F^{NK}(x,z)  
\biggl( iD_K d^\dagger_{M}(x,z) d_{N}(x,z) 
- d^\dagger_{M}(x,z) i D_K d_{N}(x,z) + \mathrm{H.c.} 
\biggr)  \nonumber\\
&+& 
d^\dagger_{M}(x,z) \, \Big(\mu^2 +  U(z) \Big) \, d^M(x,z)
\biggr]  \,, 
\en 
where 
$A(z) = \log(R/z)$, $F^{MN}(x,z) = \partial^M V^N(x,z) - 
\partial^N V^M(x,z)$ is the stress tensor of the vector field $V^M(x,z)$,   
$D^M = \partial^M - ieV^M(x,z)$ is the covariant derivative, 
$\mu^2 R^2 = (\Delta - 1) (\Delta - 3)$ is the five-dimensional mass; 
$R$ is the AdS radius, $\varphi(z) = \kappa^2 z^2$ is the 
background dilaton field;  
$\Delta = \tau + L$ is the dimension of the $d^M(x,z)$ field; 
$L$ is the orbital angular momentum, and $M_d$ is the deuteron mass. 
$U(z)$ is the confinement potential with
\eq
U(z)  =  \frac{\varphi(z)}{R^2} \, U_0 \, ,
\en
where the constant $U_0$ is fixed by the value of the deuteron
mass. In the following we work in the axial gauge for both 
vector fields $d^z(x,z) = 0$ and $V^z(x,z) = 0$. In our consideration 
we have two free parameters: $\kappa$ and $U_0$ (the latter only relevant 
for the description of the deuteron mass). 
As it will be shown later, the parameters $c_2$ and $c_3$ are constrained 
by normalization of the deuteron electromagnetic form factors. 

First we perform a Kaluza-Klein (KK) decomposition for 
the vector AdS field dual to the deuteron 
\eq\label{KK_decomposition} 
d^\mu(x,z) 
= \exp\Big[\frac{\varphi(z)-A(z)}{2}\Big] \, 
\sum\limits_n d^\mu_{n}(x) \Phi_{n}(z) \, , 
\en 
where $d^\mu_{n}(x)$ is the tower of the KK fields dual to 
the deuteron fields with radial quantum number $n$ and twist-dimension 
$\tau = 6$, and $\Phi_{n}(z)$ are their bulk profiles. 

Then we derive the Schr\"odinger-type equation of motion (EOM) for 
the bulk profile $\Phi_{n}(z)$ with
\eq 
\biggl[ - \frac{d^2}{dz^2} + \frac{4(L + 4)^2 - 1}{4z^2} 
+ \kappa^4 z^2 + \kappa^2 U_0 \biggr] \Phi_{n}(z) = 
M_{d,n}^2 \Phi_{n}(z) \;.
\en 
The analytical solutions of this EOM read 
\eq 
\Phi_{n}(z) &=& \sqrt{\frac{2n!}{(n+L+4)!}}  \, 
\kappa^{L+5} \, z^{L+9/2} \, e^{-\kappa^2 z^2/2} 
\, L_n^{L+4}(\kappa^2z^2)\,, \nonumber\\
M_{d,n}^2 
&=& 4\kappa^2 \biggl[ n + \frac{L + 5}{2} + \frac{U_0}{4}\biggr]\,,  
\en  
where $L_n^m(x)$ are the generalized Laguerre polynomials. 
Restricting to the ground state $(n=0, \ L=0)$ we get 
$M_d = 2 \kappa \, \sqrt{\frac{5}{2} + \frac{U_0}{4}}$.  
Using the central value of data for the deuteron mass $M_d = 1.875613$ GeV 
and $\kappa = 190$ MeV (fitted from data on 
electromagnetic deuteron form factors), we fix $U_0 = 87.4494$. 
We can compare this value for the deuteron scale parameter
to the analogous one of $\kappa_N$ defining the nucleon 
properties - mass and electromagnetic form factors. 
In Ref.~\cite{AdSQCD} we fixed the value to $\kappa_N \simeq 380$ MeV, 
which is 2 times bigger than the deuteron scale parameter $\kappa$. 
The difference between the nucleon and deuteron scale parameters 
can be related to the change of size of the hadronic systems -  
the deuteron as a two-nucleon bound state is 
2 times larger than the nucleon. 

In the case of the vector field dual to the electromagnetic field 
we perform a Fourier transform with respect to the Minkowski coordinate 
\eq\label{V_Fourier} 
V_\mu(x,z) = \int \frac{d^4q}{(2\pi)^4} e^{-iqx} V_\mu(q) V(q,z) 
\en 
where $V(q,z)$ is its bulk profile obeying the following EOM:  
\eq 
\partial_z \biggl( \frac{e^{-\varphi(z)}}{z} \, 
\partial_z V(q,z)\biggr) + q^2 \frac{e^{-\varphi(z)}}{z} \, V(q,z) = 0 \;.
\en 
Its analytical solution~\cite{AdSQCD_int1} can be written in the form of 
an integral representation 
introduced in Ref.~\cite{Grigoryan:2007my},  
\eq
\label{VInt}
V(Q,z) = 
\kappa^2 z^2 \int\limits_0^1 \frac{dx}{(1-x)^2} \, 
e^{-\kappa^2z^2 x/(1-x)} \, x^a \,, \quad 
a = \frac{Q^2}{4\kappa^2}\,, \quad Q^2 = - q^2  \,. 
\en
The gauge-invariant matrix element describing the interaction of the deuteron 
with the external vector field (dual to the electromagnetic field) reads 
\eq\label{M_inv} 
M_{\rm inv}^\mu(p,p') &=& - 
\biggl( G_1(Q^2) \epsilon^\ast(p') \cdot \epsilon(p) -  
\frac{G_3(Q^2)}{2M_d^2} \, \epsilon^\ast(p') \cdot q \, 
\epsilon(p) \cdot q \biggr) \, (p+p')^\mu \nonumber\\
&-& G_2(Q^2) \, 
\biggl( \epsilon^\mu(p) \, \epsilon^{\ast}(p') \cdot q 
- \epsilon^{\ast\mu}(p') \, \epsilon(p) \cdot q \biggr) 
\en 
where  $\epsilon$($\epsilon^\ast$) and
$p(p^\prime)$ are the polarization and four-momentum of the initial (final)
deuteron, and $q=p^\prime - p$ is the momentum transfer.
The three EM form factors $G_{1,2,3}$ of the deuteron
are related to the charge $G_C$, quadrupole $G_Q$ and magnetic $G_M$
form factors by
\eq
G_C = G_1+\frac{2}{3}\tau_d G_Q\,, \hspace*{.25cm}
G_M \ = \ G_2 \,,            \hspace*{.25cm}
G_Q = G_1-G_2+(1+\tau_d)G_3,   \hspace*{.25cm}
\tau_d=\frac{Q^2}{4M_d^2} \,.
\en
These form factors are normalized at zero recoil as
\eq
G_C(0)=1\,, \ \ \
G_Q(0)=M_d^2{\cal Q}_d=25.83\,, \ \ \
G_M(0)=\frac{M_d}{M_N}\mu_d=1.714 \, ,
\en
where $M_d$ and $M_N$ 
are deuteron and nucleon masses, and 
${\cal Q}_d = 7.3424$ GeV$^{-2}$ and $\mu_d = 0.8574$ 
are the quadrupole and magnetic moments of the deuteron. 
Since the deuteron is a spin-1 particle it has three EM form factors
in the one-photon-exchange approximation, due to current 
conservation and the $P$ and $C$ invariance of the EM interaction. 

We illustrate the algorithm for calculating the deuteron form factors, 
considering a particular case of the form factor 
$G_1(Q^2)$, which is generated by the second term 
in the effective action~(\ref{Eff_action}),    
\eq\label{S1} 
S^{(1)} = \int d^4xdz \, e^{-\varphi(z)} \, 
e V_\mu(x,z) \, \Big( 
i\partial^\mu d^\dagger_{\nu}(x,z) d^\nu(x,z) - 
d^\dagger_{\nu}(x,z) i\partial^\mu d^\nu(x,z) \Big) \,. 
\en 
Next we use the Kaluza-Klein decomposition~(\ref{KK_decomposition}) 
for the five-dimensional fields $d_\nu(x,z)$ 
and $d_\nu^\dagger(x,z)$ 
(restricting to the contribution of the ground states with $n=0$) 
and perform the Fourier transform for $d_\nu(x)$, $d_\nu^\dagger(x)$,   
\eq\label{dn_Fourier}
d_\nu(x) 
=  \int \frac{d^4p}{(2\pi)^4} e^{-ipx} \, \epsilon_\nu(p) \,, 
\quad\quad 
d_\nu^\dagger(x) =  \int \frac{d^4p'}{(2\pi)^4} e^{ip'x} \, 
\epsilon^\ast_\nu(p') 
\en 
and $V_\mu(x,z)$ [see Eq.~(\ref{V_Fourier})].  
Substituting expressions~(\ref{V_Fourier}) and (\ref{dn_Fourier}) 
in action (\ref{S1}) and integrating over $x$ and $z$, we get 
\eq 
S^{(1)} = (2\pi)^4 \, \int \frac{d^4p}{(2\pi)^4} \int \frac{d^4p'}{(2\pi)^4} 
\int \frac{d^4q}{(2\pi)^4} \, \delta^4(p+q-p') \, 
e V_\mu(q) \, M_{\mathrm{inv}}^{\mu, (1)}(p,p')
\en 
where $M_{\mathrm{inv}}^{\mu, (1)}(p,p')$ is part of the invariant 
matrix element of the $d + \gamma \to d$ transition containing the 
contribution of the form factor $G_1(Q^2)$ 
\eq 
M_{\mathrm{inv}}^{\mu, (1)}(p,p') = - (p+p')^\mu \, 
\epsilon^\ast(p') \cdot \epsilon(p) \,  G_1(Q^2) \,. 
\en 
In our approach the deuteron form factor $G_1(Q^2) = F(Q^2)$, 
where $F(Q^2)$ is the twist-6 hadronic form factor, 
which is given by the overlap of the square of the bulk profile dual to
the deuteron wave function (twist-6 hadronic wave function)
and the confined electromagnetic current
\eq\label{FQ2} 
F(Q^2) = \int\limits_0^\infty dz \, \Phi^2_0(z) \, 
V(Q,z) = \frac{\Gamma(6) \, \Gamma(a+1)}{\Gamma(a+6)} 
\en  
where $a = Q^2/(4 \kappa^2)$. This formula follows from the general 
and universal formula for the hadronic form factor with twist $\tau$ 
derived in Ref.~\cite{AdSQCD_int3} in terms 
of the bulk profile $\phi_\tau(z) = \sqrt{\frac{2}{(\tau-2)!}} \, 
\kappa^{\tau-1} z^{\tau-3/2} e^{-\kappa^2 z^2/2}$ dual to the 
hadronic wave function with twist $\tau$: 
\eq\label{FQ2_tau} 
F_\tau(Q^2) = \int\limits_0^\infty dz \, \phi^2_\tau(z) \, 
V(Q,z) = \frac{\Gamma(\tau) \, \Gamma(a+1)}{\Gamma(a+\tau)} 
\en   
Therefore, Eq.~(\ref{FQ2}) is the particular case of Eq.~(\ref{FQ2_tau}) 
for $\tau=6$. 

By analogy we calculate the other two deuteron form factors $G_2$ and $G_3$, 
which are expressed in terms of the same universal 
factor $F(Q^2)$: 
\eq 
G_i(Q^2) = c_i F(Q^2) \,, \quad i = 2, 3 \,.  
\en  
The parameters $c_2$ and $c_3$ are defined by 
normalization of the deuteron form factors as: 
\eq 
c_2 = G_M(0) = 1.714\,, \quad c_3 = G_M(0) + G_Q(0) - 1 = 26.544\,. 
\en 
Note that the form factor $F(Q^2)$ has the correct power-scaling 
$F(Q^2) \sim 1/(Q^2)^5$ at large $Q^2 \to \infty$.  
It can also be written in the  Brodsky-Ji-Lepage form 
derived within perturbative QCD. The deuteron form factor
is factorized in terms of the nucleon form factor 
$F_N(Q^2/4)$ and the so-called ``reduced'' nuclear form factor 
$f_d(Q^2)$~\cite{Brodsky:1983vf}: 
$F_d(Q^2) = f_d(Q^2) \, F_N^2(Q^2/4)$.  
Our result reads 
\eq 
F_d(Q^2) \equiv F(Q^2) = 
 \frac{\Gamma(6) \, \Gamma(a+1)}{\Gamma(a+6)} 
= \frac{5!}{(a+1) \ldots (a+5)} = 
f_d(Q^2) \, F_N^2(Q^2/4) 
\en 
where our predictions for $f_d(Q^2)$ and $F_N(Q^2/4)$ are 
\eq 
f_d(Q^2) = \frac{30 (a+1) (a+2)}{(a+3) (a+4) (a+5)}\,, \quad 
F_N(Q^2/4) = \frac{2}{(a+1) (a+2)}
\en 
where $a = Q^2/(4\kappa^2)$. 
Our results for the charge $G_C(Q^2)$, quadrupole $G_Q(Q^2)$ 
and magnetic $G_M(Q^2)$ form factors are shown in Figs.1-3.  
The shaded band corresponds to values of the scale parameter $\kappa$ 
in the range of 150 MeV $ < \kappa < $ 250 MeV. An increase of the 
parameter $\kappa$ leads to an enhancement of the form factors. 
The best description of the data on the deuteron form factors is obtained 
for $\kappa = 190$ MeV and is shown by the solid line. 
Data points are taken from 
Refs.~\cite{Abbott:2000ak,Holt:2012gg}. To quantify the quality of 
the fit with $\kappa = 190$ MeV we indicate the $\chi^2$ values for 
the three deuteron form factors: 
$\chi^2 = 0.2$ for ($G_C$), 
$\chi^2 = 13.8$ for ($G_Q$) and 
$\chi^2 = 2.3$ for ($G_M$). 
We would also like to 
point out that with $\kappa = 190$ MeV our result for the deuteron 
charge radius $r_C = (-6 dG_C(Q^2)/dQ^2|_{Q^2=0})^{1/2} = 
\sqrt{\frac{137}{40 \kappa^2} - {\cal Q}_d} = 1.846$~fm 
compares well with data, $r_C = 2.130 \pm 0.010$~fm~\cite{Garcon:2001sz}. 

In conclusion we stress again the main result of this paper. 
Using the soft-wall AdS/QCD model we calculate the deuteron 
electromagnetic form factors, which are given by analytical 
expressions in terms of a universal twist-6 form factor $F(Q^2)$ 
relevant for the deuteron --- hadronic system with six partons. 
Our framework gives a description of the deuteron in terms of two 
free parameters --- the dimensional parameter $\kappa$ and the confinement 
parameter $U_0$. The parameter $\kappa$ is fixed by the scale of 
the deuteron form factors and the parameter $U_0$ is fixed through 
the deuteron mass. 

\vspace*{.1cm}

\begin{acknowledgments}

This work was supported 
by the German Bundesministerium f\"ur Bildung und Forschung (BMBF)
under Grant No. 05P12VTCTG, by Marie Curie Reintegration Grant IRG 256574,
by CONICYT (Chile) Research Project No. 80140097,
by CONICYT (Chile) under Grant No. 7912010025,
by FONDECYT (Chile) under Grants No. 1140390 and No. 1141280,
and by Tomsk State University
Competitiveness Improvement Program.
V.E.L. would like to thank Departamento de F\'\i sica y Centro
Cient\'\i fico Tecnol\'ogico de Valpara\'\i so (CCTVal), Universidad
T\'ecnica Federico Santa Mar\'\i a, Valpara\'\i so, Chile and 
Instituto de F\'isica y Astronom\'ia, Centro de Astrof\'isica de Valpara\'iso, 
Universidad de Valpara\'iso, Chile for warm hospitality.

\end{acknowledgments}

\begin{figure}
\begin{center}

\epsfig{figure=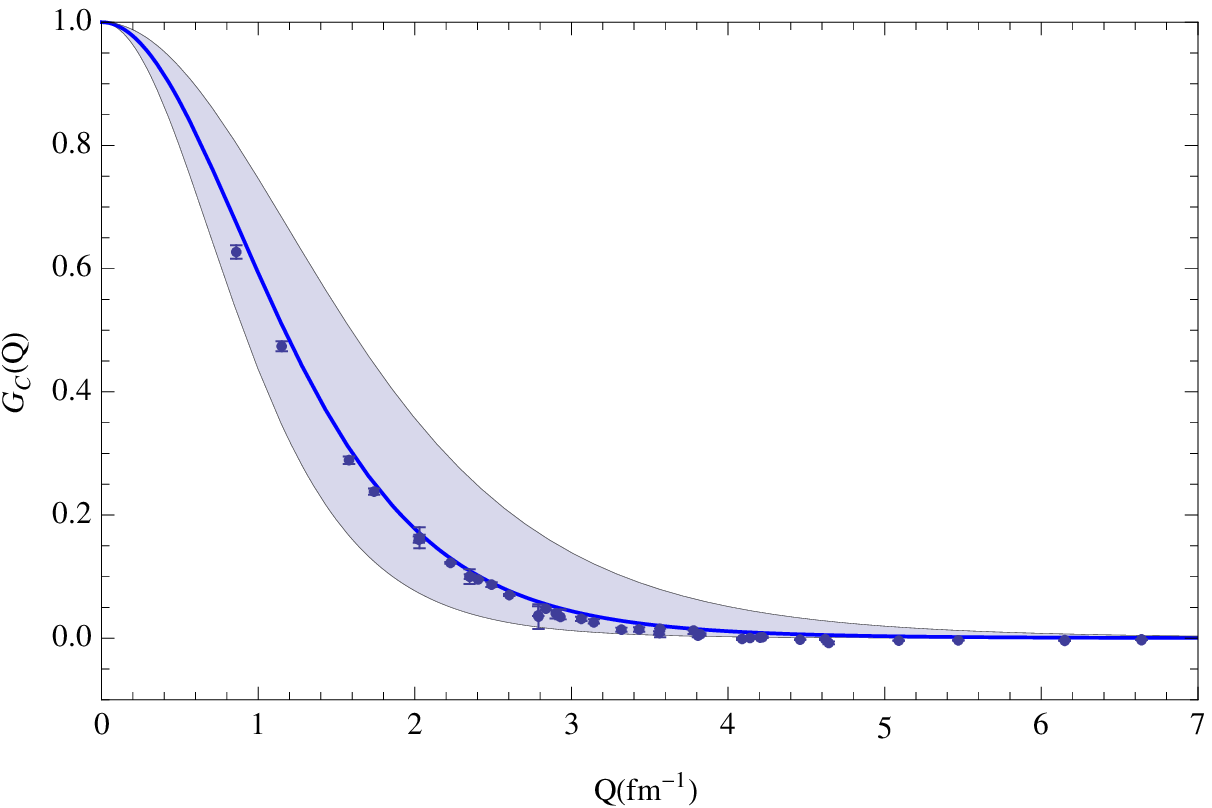,scale=.75}
\noindent
\caption{Charge deuteron form factor.}

\vspace*{.5cm}

\epsfig{figure=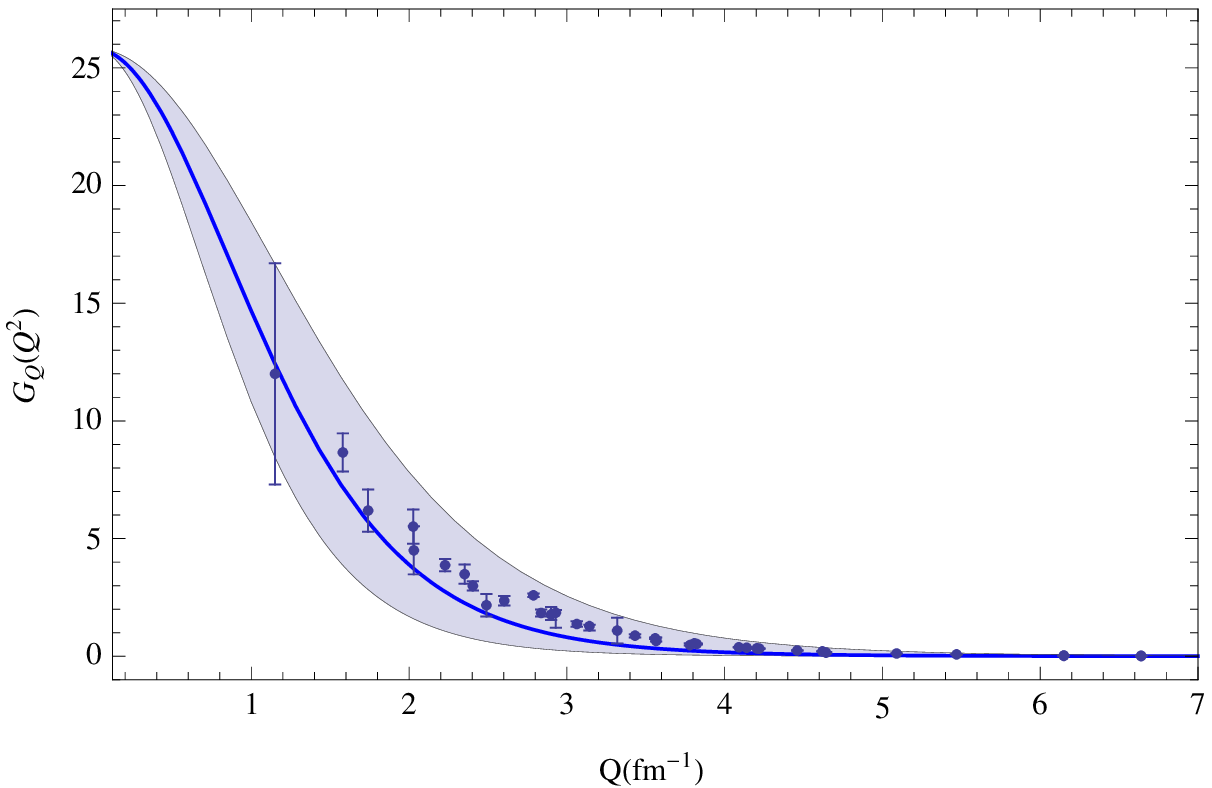,scale=.75}
\noindent
\caption{Quadrupole deuteron form factor.}

\vspace*{.5cm}

\epsfig{figure=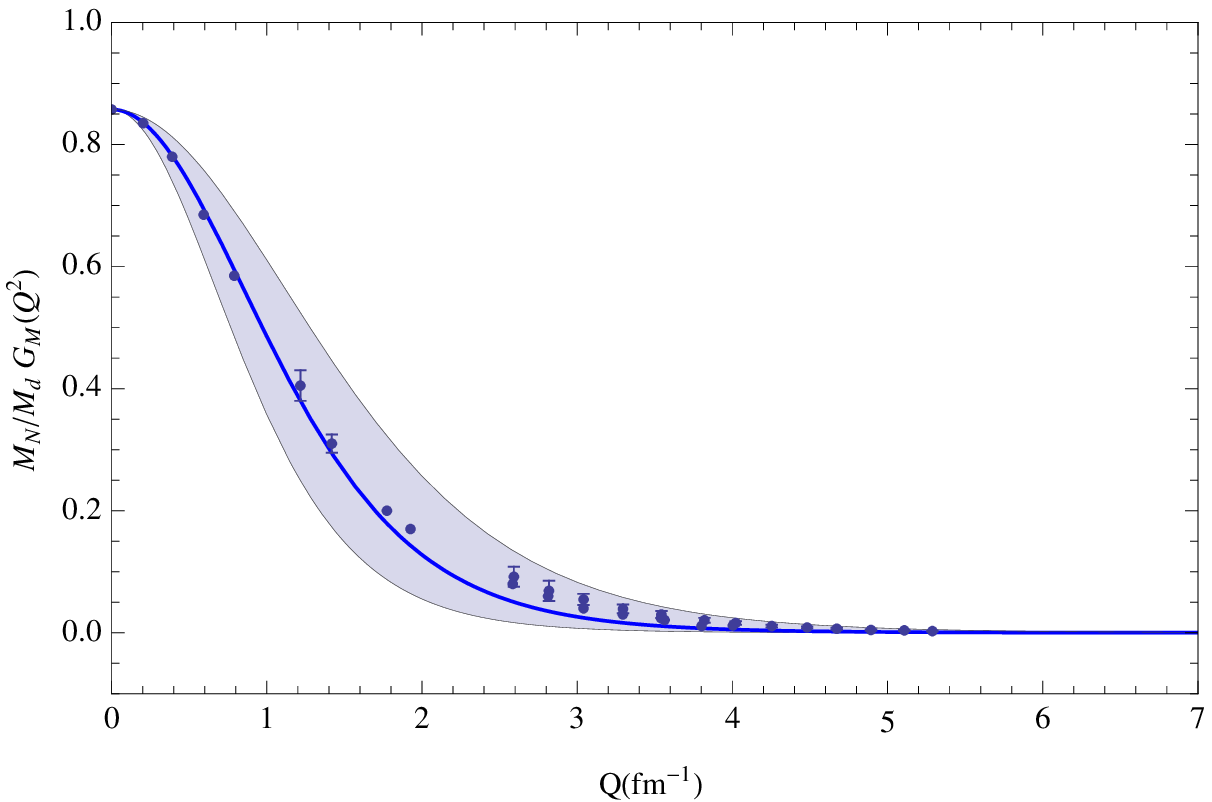,scale=.75}
\noindent
\caption{Magnetic deuteron form factor.}

\end{center}
\end{figure}

\end{document}